\documentclass[aps,prb,preprint,showpacs,superscriptaddress]{revtex4-1}
\usepackage{graphicx}
\usepackage{amsmath}

\begin{document}

\title{Influence of Intrinsic Electronic Properties on Light Transmission through Subwavelength Holes on Gold and MgB$_{2}$ Thin Films}

\author{Xu Fang}
\affiliation{Institute of Physics and Beijing National Laboratory for Condensed Matter Physics, Chinese Academy of Sciences, Beijing 100190, China}
\author{Chenggang Zhuang}
\affiliation{State Key Laboratory for Artificial Microstructure and Mesoscopic Physics, School of Physics, Peking University, Beijing 100871, China}
\affiliation{Department of Physics, Temple University, Philadelphia, Pennsylvania 19122, USA}
\author{Zhenchao Wen}
\affiliation{Institute of Physics and Beijing National Laboratory for Condensed Matter Physics, Chinese Academy of Sciences, Beijing 100190, China}
\author{Xiufeng Han}
\affiliation{Institute of Physics and Beijing National Laboratory for Condensed Matter Physics, Chinese Academy of Sciences, Beijing 100190, China}
\author{Qingrong Feng}
\affiliation{State Key Laboratory for Artificial Microstructure and Mesoscopic Physics, School of Physics, Peking University, Beijing 100871, China}
\author{Xiaoxing Xi}
\affiliation{Department of Physics, Temple University, Philadelphia, Pennsylvania 19122, USA}
\author{Franco Nori}
\affiliation{Advanced Science Institute, RIKEN, Wako-shi, Saitama, 351-0198, Japan}
\affiliation{Department of Physics, University of Michigan, Ann Arbor, Michigan 48109, USA}
\author{Xincheng Xie}
\affiliation{Institute of Physics and Beijing National Laboratory for Condensed Matter Physics, Chinese Academy of Sciences, Beijing 100190, China}
\affiliation{International Center for Quantum Materials, Peking University, Beijing 100871, China}
\author{Qian Niu}
\affiliation{International Center for Quantum Materials, Peking University, Beijing 100871, China}
\affiliation{Department of Physics, University of Texas, Austin, Texas 78712, USA}
\author{Xianggang Qiu}
\email[]{xgqiu@aphy.iphy.ac.cn}
\affiliation{Institute of Physics and Beijing National Laboratory for Condensed Matter Physics, Chinese Academy of Sciences, Beijing 100190, China}

\date{\today}

\begin{abstract}
We show how intrinsic material properties modify light transmission through subwavelength hole arrays on thin metallic films in the THz regime. We compare the temperature-dependent transmittance of Au films and MgB$_{2}$ films. The experimental data is consistent with analytical calculations, and is attributed to the temperature change of the conductivity of both films. The transmission versus conductivity is interpreted within the open resonator model when taking the skin depth into consideration. We also show that the efficiency of this temperature control depends on the ratio of the transmission peak frequency to the superconducting energy gap in MgB$_{2}$ films.
\end{abstract}

\pacs{73.20.Mf,78.67.-n,74.78.-w,42.25.Bs}

\maketitle

\section{introduction}
Light transmission through subwavelength holes on metallic films works beyond the traditional diffraction limit when light excites evanescent surface waves on the film \cite{Ebbesen1998}. This so-called Extraordinary Optical Transmission (EOT) can play an important role in miniaturizing opto-electronic devices \cite{Ozbay2006}. The frequency and intensity of the EOT peaks depend on how evanescent surface waves are excited by the incident light, which is determined by the geometry of the holes, as well as the intrinsic properties of the metal and surrounding dielectrics \cite{Vidal2010}. Although geometric factors of the excitation have been extensively investigated, the influence of the intrinsic properties of the metal remains to be explored. Especially in the THz regime, metals are routinely treated as perfect electromagnetic conductors, completely overlooking their specific material properties. Differences of the EOT between good and poor conductors have been reported by comparing samples with the same structure but made from different materials \cite{Rodrigo2008,Przybilla2006,Azad2006}. However, this comparison can be easily affected by the variation of both film thickness and hole diameter in different samples. We overcome such difficulties by investigating the temperature-dependent EOT, taking advantage of a nearly continuous variation of material properties through a fine control of the temperature. Besides that, we compare  MgB$_{2}$ (a low-temperature superconductor) with Au (a noble metal), whose complex conductivities change with temperature differently. By investigating the similarities and differences of the evolutions of the EOT peaks on these two films, we demonstrate \textit{the influence of intrinsic material properties on EOT in the THz regime}.

An outcome of this study, for future possible applications of subwavelength optics, is that \textit{the EOT on metallic films can be tuned effectively by just varying the temperature}. Previously, temperature control of EOT was realized either by changing the properties of dielectric surroundings \cite{Isaac2009}, or by using thin films made of doped semiconductors \cite{Rivas2004}. Here we report a $\sim 30\%$ increase in the maximum transmittance of a gold film when the temperature decreases from 300 K to 5 K, which is attributed to the change of the properties of gold with temperature. Refs. \onlinecite{Tsiatmas2010,Tian2010} observe EOT on thin films of YBa$_{2}$Cu$_{3}$O$_{7}$, a high-temperature superconductor. Here we use MgB$_{2}$, a low-temperature superconductor, whose low temperature properties are described more accurately by using BCS theory. We observe that both the intensity and the shape of the EOT peaks change when MgB$_{2}$ enters the superconducting phase. Through a comparison of EOT peaks at different frequencies, we explore how the interplay between superconductivity and EOT depends on both the temperature and the frequency.

\section{Sample preparation and measurement}

\begin{figure}
\includegraphics[width=8.3cm]{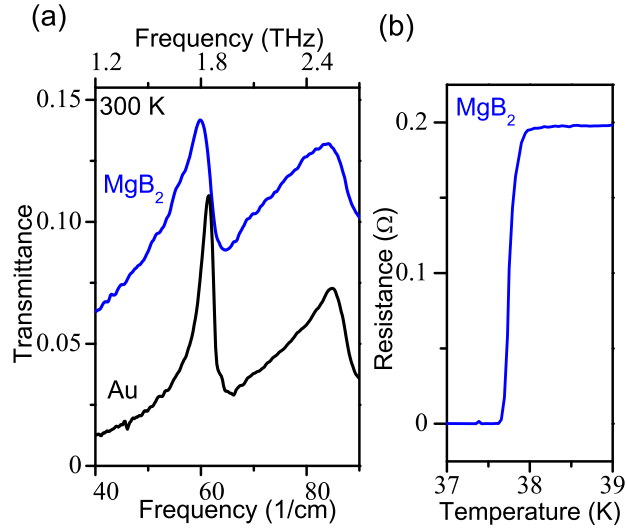}
\caption{\label{fig1} (Color online) Experimental results for a MgB$_{2}$ film and a Au film with arrays of subwavelength holes. (a) Transmission spectra for both films at 300 K. (b) Resistance versus temperature for the MgB$_{2}$ film.}
\end{figure}

We grew c-axis-oriented MgB$_{2}$ films \cite{Zeng2002}, and also Au films, on MgO substrates. Wedged [111] MgO substrates were used to avoid the interference of light between the front and back surfaces of each substrate. The $c$-axis oriented MgB$_{2}$ films were grown by the hybrid physical-chemical vapor deposition technique. The Au film was grown by magnetron sputtering. Both the MgB$_{2}$ and Au films were 50 nm thick, which was checked by measuring their cross-section thickness under a scanning electron microscope. Ultraviolet lithography and reactive ionic etching were used for patterning the arrays of holes. The superconducting transition of the fabricated MgB$_{2}$ film occurred at 37.6 K, with a transition width of 0.3 K [Fig. \ref{fig1}(b)]. Meanwhile, our gold films showed the typical behavior of a normal metal in the entire temperature range. Each pattern was a square array of circular holes, with a total effective area of $10\times10$  mm$^{2}$.

Samples and a bare MgO substrate were put into an optical cryostat. Temperature was varied from 300 K to 5 K. The transmittance spectra of the samples and the substrate were taken in a Fourier transform infrared spectrometer (ABB Bomem DA8). The light had a normal incidence, with its beam diameter limited, by a diaphragm, to 5 mm. The transmittance of the samples was obtained by subtracting the spectra of the substrate from the spectra of the samples.

\section{Results and discussion}

Figure \ref{fig1}(a) shows the transmittance of a MgB film and a gold film at 300 K. The lattice constant of the hole array of both samples is 50 $\mu$m and the hole diameter is 25 $\mu$m. Transmission peaks at approximately 60 cm$^{-1}$ (or 1.8 THz) and 85 cm$^{-1}$ (or 2.5 THz) are observed in both samples. These are the EOT peaks because their corresponding wavelengths are far larger than the hole diameter \cite{Abajo2007}, and can be assigned to the (1,0) and (1,1) modes of the Metal/MgO interface, respectively \cite{Ghaemi1998}.

\begin{figure}
\includegraphics[width=8.3cm]{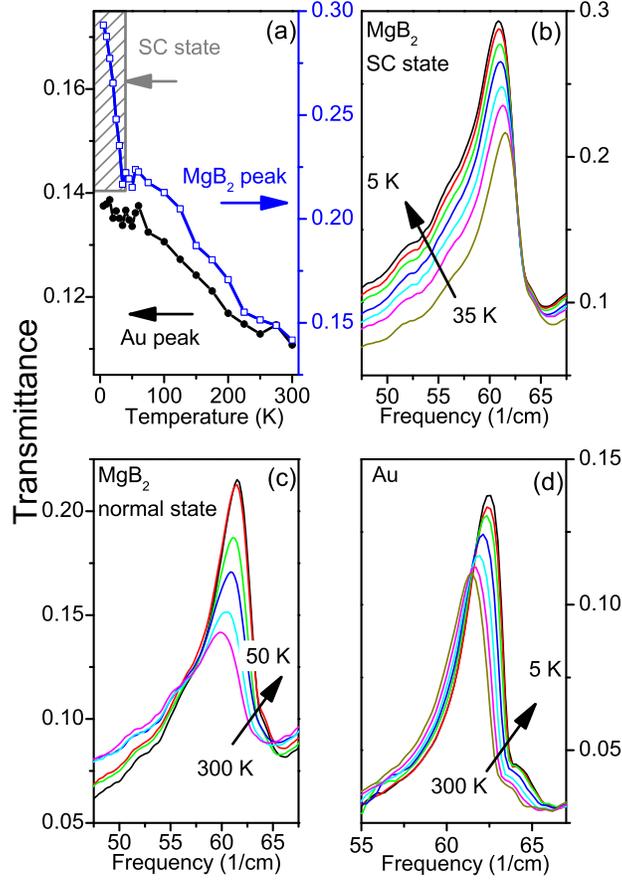}
\caption{\label{fig2} (Color online) Experimental spectra of MgB$_{2}$ and Au around 60 cm$^{-1}$ at different temperatures. (a) The maxima of the transmittance at different temperatures. The hatched area shows the superconducting (SC) regime of the MgB$_{2}$ film. (b) Transmittance spectra of the MgB$_{2}$ film in the SC state. Temperatures change from 5 K to 35 K, with a step of 5 K. (c) Representative spectra of the MgB$_{2}$ film in the normal state. Temperatures change from 50 K to 300 K, with a step of 50 K. (d) Representative spectra of the Au film. Temperatures are 5 K, and also from 50 K to 300 K with a step of 50 K.}
\end{figure}

Light transmission spectra at 300 K show the combined effects of the geometry of the array of holes, as well as the intrinsic properties of the metallic and dielectric materials. Indeed, \textit{varying the temperature highlights the effect of the material properties}. Figure \ref{fig2} shows how the spectra around 60 cm$^{-1}$ change when decreasing the temperature. The spectral peak of the Au film becomes narrower and shows a small blue shift, and the peak intensity increases smoothly [Figs. \ref{fig2}(a) and \ref{fig2}(d)]. Such a trend continues down to the lowest measured temperature of 5 K. For the MgB$_{2}$ film, the spectra show similar behaviors as those of the Au film in the normal state [Fig. \ref{fig2}(c)]. However, in the superconducting state, the spectra exhibit qualitatively different behaviors from the normal state. In the superconducting state [Fig. \ref{fig2}(b)], the peak becomes slightly \textit{broader} (instead of \textit{narrower}) and shows a very slight \textit{red} shift, instead of the \textit{blue} shift observed before, as the temperature is decreased. Moreover, the spectra show a Fano line-shape \cite{Genet2003} with the high-frequency side of the peak being nearly independent of the temperature and the low-frequency side becoming broader at low temperatures, resulting in a very slight increase of the asymmetry of the Fano line-shape. This very small increase in the asymmetry could be the result of the increase in the available low-lying quasi-particle states, which act as the continuum channel of the Fano resonance. The maximum transmittance increases with decreasing temperature in the entire temperature range, with a much faster increase rate in the superconducting state than in the normal state [Fig. \ref{fig2}(a)]. For both MgB$_{2}$ and Au films, the effect of changing temperature is much more pronounced on the peak intensity than on the peak frequency. Because the EOT peak frequency and intensity correspond to the eigen-mode and magnitude of evanescent surface waves, respectively, we conclude that, the eigen-modes of evanescent surface waves are slightly modified, while their amplitudes increase significantly with decreasing temperature.

\begin{figure}
\includegraphics[width=8.3cm]{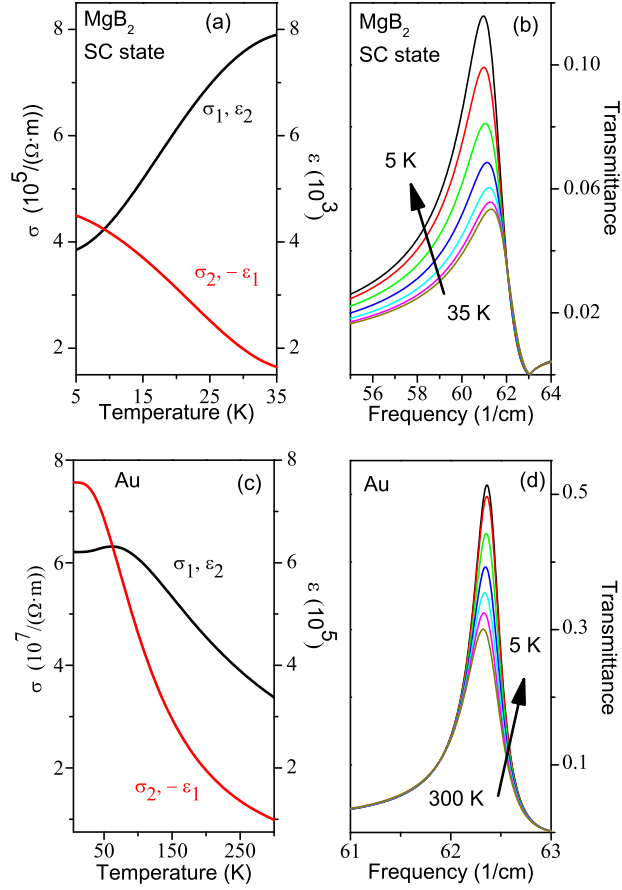}
\caption{\label{fig3} (Color online) Analytically calculated results of the conductivity $\sigma$, the relative permittivity $\epsilon$, and transmittance spectra around 60 cm$^{-1}$. (a) The conductivity $\sigma$ and the relative permittivity $\epsilon$ of MgB$_{2}$ at 60 cm$^{-1}$, when MgB$_{2}$ is in the superconducting (SC) state.  (b) Transmittance of the SC MgB$_{2}$ film. Temperatures change from 5 K to 35 K, with a step of 5 K. (c) The conductivity $\sigma$ and the relative permittivity $\epsilon$ of Au at 60 cm$^{-1}$. (d) Transmittance of the Au film with temperatures at 5 K, and also from 50 K to 300 K with a step of 50 K.}
\end{figure}

The spectral line-shape and their temperature dependence shown above can be described through the change of the complex conductivity $\sigma$ $(= \sigma_{1} + i\sigma_{2})$ of MgB$_{2}$ and Au. Because MgB$_{2}$ is a conventional BCS superconductor, its complex conductivity in the superconducting state can be calculated using BCS theory \cite{Mattis1958,Kaindl2002}. The obtained temperature-dependent values of $\sigma_{\text{MgB}_{2}}$ at 60 cm$^{-1}$ are shown in Fig. \ref{fig3}(a). In the superconducting state, with decreasing temperature $\sigma_{1}$ decreases while $\sigma_{2}$ increases, which reflects the opening of the superconducting gap and the depletion of thermally-excited quasiparticles when the temperature decreases. Meanwhile, we calculate $\sigma_{\text{Au}}$ using an extended Drude model \cite{Ozdemir2003}, and the calculated values at 60 cm$^{-1}$ are shown in Fig. \ref{fig3}(c). For Au, in contrast to MgB$_{2}$ in the superconducting state, both $\sigma_{1}$ and $\sigma_{2}$ increase with decreasing temperature, due to the increase in electron relaxation time coming from the suppression of electron-phonon scattering at low temperatures. The complex relative permittivity $\epsilon$ $(= \epsilon_{1} + i\epsilon_{2})$ of both materials is also shown in Figs. \ref{fig3}(a) and \ref{fig3}(c), which is obtained through the equation $\epsilon = i \sigma/(\epsilon_{0} \omega)$, where $\epsilon_{0}$ is the permittivity of free space, and $\omega$ is the angular frequency. We stress there are \textit{no adjustable parameters} in the above calculations (see Appendix \ref{AppendixA}). The transmission spectra at different temperatures are then obtained by analytically solving Maxwell's equations \cite{Huang2007} using the obtained $\sigma_{\text{MgB}_{2}}$ and $\sigma_{\text{Au}}$, with representative spectra shown in Figs. \ref{fig3}(b) and \ref{fig3}(d). The film thickness and lattice constant are experimental values. Square holes are adopted in the model \cite{Huang2007}, with its side length as 5 $\mu$m. The area of the holes in the calculation is smaller than the experimental value, which may be due to the difference in the cut-off wavelength and the electromagnetic field distribution between square holes (in the calculations) and circular holes (in the experiments). The temperature-dependent features, such as the spectral line-shape and intensity, of the EOT peaks on both films are reproduced (Figs. \ref{fig2} and \ref{fig3}).

Here we use relatively thin films, so the light may transmit through the unperforated part of the film. We note that several previous reports on light transmission through artificial structures on superconducting thin films did not consider this direct transmission. The transmittance of a plain superconducting film changes significantly with temperature in the frequency range below and near the superconducting energy gap. The temperature-dependent characteristics of this direct transmission can mix up with those of the EOT. This can be a serious problem when the film thickness is less than the skin depth. Without considering and excluding the possibility of direct transmission, investigations on the optical properties of superconducting films with artificial structures are prone to error. In our work, the possible influence of the direct transmission on the EOT has been excluded because: 1) the high-frequency and the low-frequency sides of the spectra show different temperature dependences (Fig. \ref{fig2}(b)); 2) these temperature-dependent characteristics are reproduced by the analytical solution  (Fig. \ref{fig3}(b)) where the direct transmission is assumed to be zero in amplitude. So we conclude that the variation in spectra reveals the change of light transmission due to EOT, which originates from the temperature dependence of the intrinsic properties of MgB$_{2}$ and Au.

We notice that the EOT peaks in MgB$_{2}$ and Au films are significantly higher at low temperatures, with their $\sigma_{1}$ change in opposite directions with decreasing temperature [Fig. \ref{fig3}]. This is quite different from reports on either doped semiconductor films in THz, or metal films at higher frequencies \cite{Vidal2010}, where an increase in $\sigma_{1}$ generally corresponds to a decrease in EOT intensity. This is because films have $\sigma_{1}\ll\sigma_{2}$ for either doped semiconductors in the THz regime, or noble metals in the visible and near-infrared regimes. So $\sigma_{1}$ can be safely treated as a small perturbation in the theoretical analysis, which describes the dissipation of evanescent surface waves, and relates with the EOT peak width and strength. In our experiment, both MgB$_{2}$ and Au films have $\sigma_{1}\approx\sigma_{2}$ . So both $\sigma_{1}$ and $\sigma_{2}$ have to be treated together to describe the THz EOT of perforated metallic films. To achieve this, we incorporate $\sigma_{1}$ and $\sigma_{2}$ into the parameters of the skin depth $\delta$, which describes the depth the electromagnetic field penetrates into metallic films. In the THz regime, where metals have $\sigma_{1},\sigma_{2}\gg0$,
\begin{equation}
\delta=c[\omega \text{Im}\sqrt{i \sigma/(\epsilon_{0} \omega)}]^{-1},
\end{equation}
where $c$ is the speed of light in vacuum \cite{Jackson1998}. Although $\sigma_{\text{MgB}_{2}}$ and $\sigma_{\text{Au}}$ have different temperature dependences, both $\delta_{\text{MgB}_{2}}$ and $\delta_{\text{Au}}$ decrease monotonously, as both materials become better electromagnetic conductors with decreasing temperature.

Holes on the films form an array of open resonators, where evanescent surface waves resonate with incident and transmitted light \cite{Bliokh2008}. The total quality factor of these resonators includes a term $Q_{\text{dis}}$, which describes the dissipation of evanescent surface waves in these resonators. For a closed resonator with highly conductive boundaries:
\begin{equation}
Q_{\text{dis}}=G/\delta,
\end{equation}
where $G$ is a geometric factor describing the shape of the cavity \cite{Jackson1998}. Because the electromagnetic field is highly confined inside subwavelength holes in the THz regime \cite{Pendry2004}, open and closed resonators should have a similar $Q_{\text{dis}}$. Because both $\delta_{\text{MgB}_{2}}$ and $\delta_{\text{Au}}$ decrease with decreasing temperature, $Q_{\text{dis}}$ increases for both films at low temperatures. This explains why at low temperatures the intensity of the EOT increases for both MgB$_{2}$ and Au films.

This qualitative insight also explains another apparent discrepancy between our experiments and calculations on metal films at higher frequencies \cite{Vidal2010}. In our experiments, the EOT peaks for both MgB$_{2}$ and Au films increase in intensity when these materials approach the prefect-conductor limit. However, the EOT intensity in noble metal films can be higher than that in perfect-conducting films in the visible and near-infrared regime, which is attributed to the finite value of the skin depth in real metals \cite{Vidal2010}. This is because in the visible and near-infrared regime, the hole diameter is usually only slightly larger than $\delta$. So a change in $\delta$ will result in a significant enlargement of the effective hole diameter and thus influencing the geometric factor $G$ in Ref. \onlinecite{Vidal2010}. However, in our experiments $G$ is nearly constant. This can be seen as follows: $\delta \approx$ 370 nm and 40 nm for the superconducting MgB$_{2}$ and Au in our experiments, respectively. Thus, our hole diameter is several orders of magnitude larger than our $\delta$; so the change in $\delta$ does not significantly influence the effective hole size and results in a nearly constant G. Thus, in our case, the change in EOT intensity is only a reflection of the change of $\delta$.

\begin{figure}
\includegraphics[width=8.3cm]{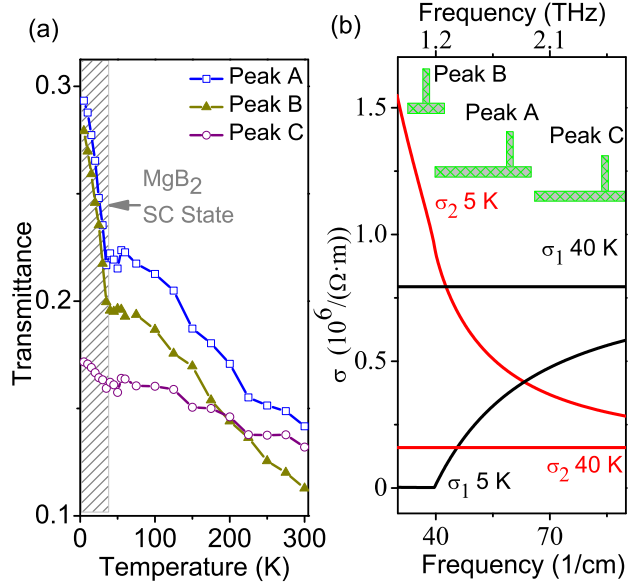}
\caption{\label{fig4} (Color online) Transmittance and conductivity of three EOT peaks on two MgB$_{2}$ films. (a) Experimental result of the maxima of the transmittance of three EOT peaks. The hatched area shows the SC regime of the MgB$_{2}$ films. (b) Results of calculations for the real $\sigma_{1}$ and imaginary $\sigma_{2}$ parts of the conductivity $\sigma$ of MgB$_{2}$ at 5 K (in the SC state) and 40 K (in the normal state). The horizontal bars at the top show the approximate spans of the three peaks. Vertical bars show the frequencies of the maximum transmittance.}
\end{figure}

Evanescent surface waves on superconducting MgB$_{2}$ films have very low dissipation, because Cooper pairs have a pure inductive response to the driven optical field at finite frequency below the superconducting energy gap \cite{Rakhmanov2010,Ortolani2008,Savelev2010}. Since photons with energy larger than the superconducting energy gap can break Cooper pairs, the effect of the superconducting transition on the EOT is expected to be much weaker when the frequency goes up. To examine this, we investigated three EOT peaks with different frequencies [Fig. \ref{fig4}]. Two of the peaks (A, C) are from the MgB$_{2}$ sample discussed above, and another peak (B) is from another MgB$_{2}$ sample with a different periodicity of the hole array. Details regarding the sample pattern and temperature-dependent spectra are available in Appendix \ref{AppendixB}. MgB$_{2}$ has two superconducting energy gaps and the lower one $[= 2\Delta_{\pi} \sim$ 43 cm$^{-1}]$ dominates its optical properties \cite{Kuzmenko2007}. These three EOT peaks correspond to the regime: slightly lower (B), slightly higher (A), and much higher (C) than $2\Delta_{\pi}$. It is clear from Fig. \ref{fig4}(a) that a phase transition is clearly revealed in the temperature-dependent transmittance maxima only for peaks with energy slightly lower and slightly higher than $2\Delta_{\pi}$. The phase transition can still be identified from the line-shape change of the spectra (see Appendix \ref{AppendixB}), but the change of transmittance maximum is much weaker for the peak (C) with energy much higher than $2\Delta_{\pi}$. Figure \ref{fig4}(b) shows the calculated conductivity of MgB$_{2}$ at 5 K (in the superconducting state) and 40 K (in the normal state) based on the Mattis-Bardeen model \cite{Mattis1958}. The difference between the conductivity at two temperatures diminishes for higher frequencies; indeed, in the visible or near-infrared regime, this difference should be very small. We also investigated EOT peaks in the mid-infrared regime (see Appendix \ref{AppendixB}), but there the superconducting transition cannot be identified in our spectra.

\section{conclusion}
In conclusion, we show that light transmission through subwavelength hole arrays on metallic thin films can be controlled by modifying the intrinsic electronic properties of the films. The real part of the conductivity alone cannot describe the strength and width of the EOT peaks in the THz regime. The temperature dependence of the EOT peaks is understood within the open resonator model, where both the real and the imaginary parts of the conductivity are taken into consideration. As we have shown, the effect of the superconducting transition on the EOT depends on the ratio of the superconducting energy gap to the frequency of the EOT. Besides the \textit{temperature control} used here, the superconducting energy gap can be easily tuned with a magnetic field or an electric current, providing additional interesting ways to control the EOT of superconducting films (see Appendix \ref{AppendixC} for more detailed discussion).

\section{acknowledgments}
This work was supported by the NSF of China (Grant No.~10974241), MOST-China (973 projects No. 2009CB929100, 2009CB930700 and 2006CD601004), and the Knowledge Innovation Program of CAS (Grant No.KJCX2-EW-W02). Also by the LPS-NSA-ARO, NSF, JSPS-FIRST and MEXT. The work at Temple University is supported by ONR under Grant No. N00014-10-1-0164.

\newpage

\newpage

\newpage

\newpage

\clearpage

\appendix

\section{\label{AppendixA} Parameters used in calculation}

\begin{table*}[h]
\caption{Parameters used to calculate the conductivity of MgB$_{2}$ in Fig. \ref{fig3}(a).}
\begin{ruledtabular}
\begin{tabular}{lrr}
\textrm{MgB$_{2}$ parameters}&
\textrm{Value}&
\textrm{Source}\\
\colrule
plasmon frequency of the normal state, $\omega_{\text{pl}}$& 1.5 eV & Ref. \onlinecite{Kaindl2002} \\
scattering rate of the normal state, $\tau^{-1}$& 37 meV & Ref. \onlinecite{Kaindl2002} \\
superconducting energy gap, $2\Delta_{0}$& 5 meV & Ref. \onlinecite{Kaindl2002} \\
transition temperature, $T_{c}$& 37.6 K & our experiment \\
\end{tabular}
\end{ruledtabular}
\end{table*}

\begin{table*}
\caption{Parameters used to calculate the conductivity of Au in Fig. \ref{fig3}(c).}
\begin{ruledtabular}
\begin{tabular}{lrr}
\textrm{Au parameters}&
\textrm{Value}&
\textrm{Source}\\
\colrule
permittivity at 60 cm$^{-1}$ at 300 K, $\epsilon$& $-9.87\times10^{4}+i 3.37\times10^{5}$ & Ref. \onlinecite{Ordal1983} \\
Fermi-surface average of scattering probability, $\Gamma$ & 0.55 & Ref. \onlinecite{Ozdemir2003} \\
thermal linear expansion coefficient, $\gamma$ & $14.2\times10^{-6} K^{-1}$ & Ref. \onlinecite{Ozdemir2003} \\
fractional Umklapp scattering, $\Delta$ & 0.77 & Ref. \onlinecite{Ozdemir2003}\\
Debye temperature, $\theta_{D}$ & 185 K & Ref. \onlinecite{Ozdemir2003}\\
Fermi energy, $E_{F}$ & 5.51 eV & Ref. \onlinecite{Ozdemir2003} \\
Poisson's number, $\mu$ & 0.42 & Ref. \onlinecite{Ozdemir2003} \\
\end{tabular}
\end{ruledtabular}
\end{table*}

\clearpage

\section{\label{AppendixB} Supplementary Figures}

\begin{figure}[h]
\includegraphics[width=8.3cm]{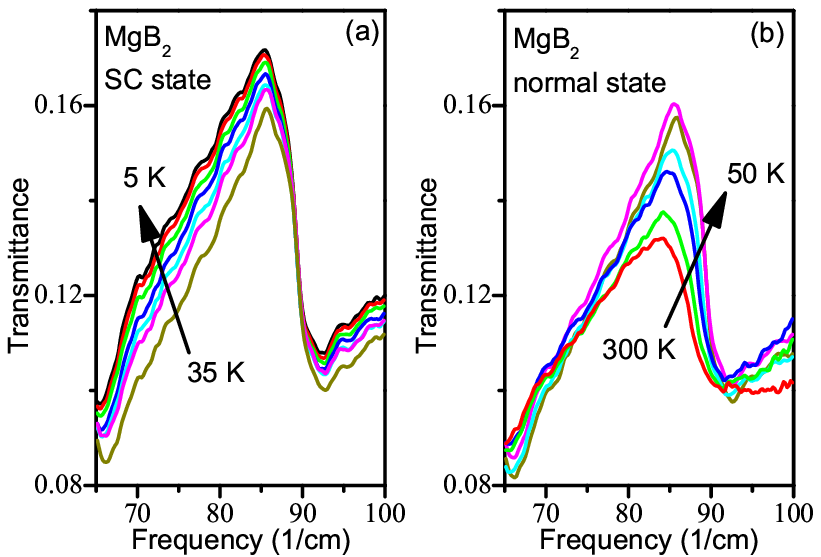}
\caption{\label{Appendixfig1} (Color online) Experimental transmittance around an EOT peak of the MgB$_{2}$ film. The film thickness is 50 nm. The array of holes has a lattice constant of 50 $\mu$m and a hole diameter of 25 $\mu$m. (A) Transmission spectra when MgB$_{2}$ is in the SC state. Temperatures change from 5 K to 35 K, with a step of 5 K. (B) Representative spectra when MgB$_{2}$ is in the normal state. Temperatures change from 50 K to 300 K, with a step of 50 K.}
\end{figure}

\clearpage

\begin{figure}
\includegraphics[width=8.3cm]{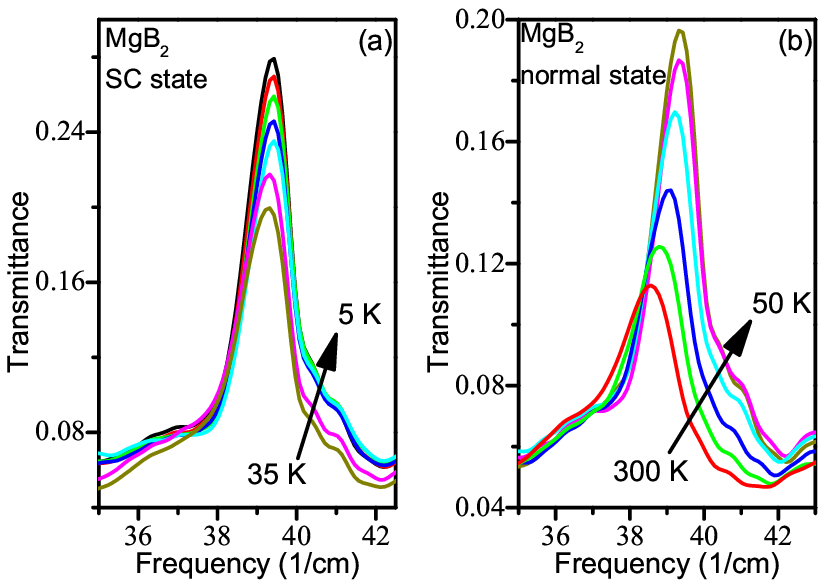}
\caption{\label{Appendixfig2} (Color online) Experimental transmittance spectra around an EOT peak on a MgB$_{2}$ film. The film thickness is 50 nm. The array of holes has a lattice constant of 80 $\mu$m and a hole diameter of 40 $\mu$m. (A) Spectra when MgB$_{2}$ is in the SC state. Temperatures change from 5 K to 35 K, with a step of 5 K. (B) Representative spectra when MgB$_{2}$ is in the normal state. Temperatures change from 50 K to 300 K, with a step of 50 K.}
\end{figure}

\clearpage

\begin{figure}
\includegraphics[width=8.3cm]{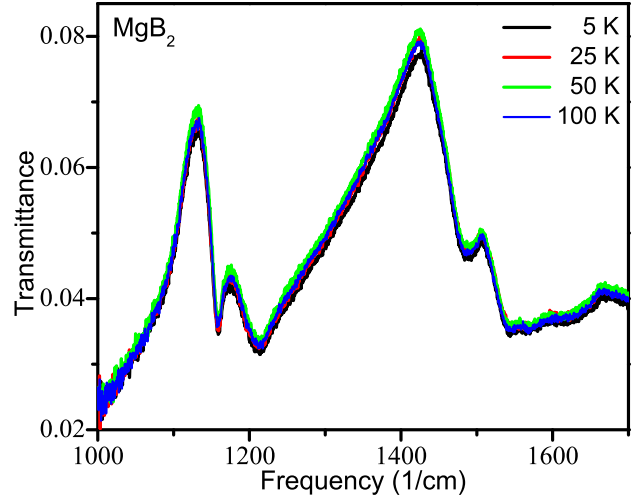}
\caption{\label{Appendixfig3}(Color online) Experimental transmittance spectra around several EOT peaks on a MgB$_{2}$ film in the mid-infrared regime. The film thickness is 50 nm. The array of holes has a lattice constant of 6 $\mu$m and a hole diameter of 3 $\mu$m. Representative spectra are shown here, with temperatures at 5 K, 25 K, 50 K, and 100 K. These are nearly overlapping.}
\end{figure}

\clearpage

\section{\label{AppendixC} Modulation of the EOT of Superconducting Films by Using Electric and Magnetic Fields}
Both static electric and magnetic fields are able to decrease the superconducting energy gap, and even turn the superconducting state into normal state when these fields are strong enough. This property can be used to modulate the intensity of the EOT of superconducting films.

As an example, we now discuss the EOT peak at 60 cm$^{-1}$ (Peak A in the main text). As shown in Fig. 2 and Fig. 4, the transmittance of this peak is 0.293 at 5 K, when MgB$_{2}$ is in the superconducting state. A strong enough electric or magnetic field is able to destroy the superconductivity of MgB$_{2}$. As a reasonable extrapolation of the data in Fig. 2 and Fig. 4, the transmittance will now decrease to a level close to that at 50 K in Fig. 2. So the maximum relative modulation amplitude at 5 K is approximately:
\begin{equation*}
\frac{\text{T(5 K, SC state)}-\text{T(50 K, normal state)}}{\text{T(50 K, normal state)}}\approx35\%,
\end{equation*}
where $T$ is the transmittance.

Sufficiently strong magnetic or electric fields are usually required to completely destroy the superconductivity of MgB$_{2}$ at low temperatures. As a type-II superconductor, the upper critical magnetic field of MgB$_{2}$ thin films is above 20 T at 5 K \cite{Zeng2002}. The zero-field critical current density is of the order of $10^{11} \text{A/m}^{2}$ at 5 K \cite{Zeng2002}. It is possible to achieve an effective modulation at lower field strength when the magnetic field and electric field are used together, because a moderate magnetic field reduces the critical current density by orders of magnitude \cite{Zeng2003}.

\clearpage

\end{document}